\documentclass[prl,twocolumn,aps,showpacs,amsmath,amssymb,superscriptaddress,floatfix,longbibliography]{revtex4-1}

\usepackage{epsfig,color}
\usepackage{graphicx}
\usepackage{dcolumn}% Align table columns on decimal point
\usepackage{bm}% bold math
\usepackage{amsmath, amsfonts, amssymb,mathrsfs}
\usepackage{pstricks}
\usepackage{amsxtra}
\usepackage{amsthm}
\usepackage{natbib}
\usepackage{qcircuit}
\usepackage{array}

\newcommand{\ket}[1]{\mbox{$|#1\rangle$}}
\newcommand{\bra}[1]{\mbox{$\langle#1|$}}

\def\be{\begin{equation}}      % with numbering
\def\ee{\end{equation}}
\def\beu{\begin{equation*}}   % without numbering
\def\eeu{\end{equation*}}

\DeclareMathOperator{\trace}{Tr}      % Trace of matrix
\providecommand{\mean}[1]{\langle#1\rangle}
\theoremstyle{definition}

\definecolor{new}{rgb}{.08,.05,.8}

\newcommand{\delete}[1]{{}} 

\begin{document}
\title{Scalable probes of measurement-induced criticality}
\date{\today}
\author{Michael J. Gullans}
\affiliation{Department of Physics, Princeton University, Princeton, New Jersey 08544, USA}
\author{David A. Huse}
\affiliation{Department of Physics, Princeton University, Princeton, New Jersey 08544, USA}
\affiliation{Institute for Advanced Study, Princeton, New Jersey 08540, USA}

\begin{abstract}
We uncover a  local order parameter for measurement-induced phase transitions: the average entropy of a single reference qubit initially entangled with the system.  
 Using this order parameter, we identify scalable probes of measurement-induced criticality (MIC) that are immediately applicable to advanced quantum computing platforms. 
 We test our proposal on a 1+1 dimensional stabilizer circuit model that can be classically simulated in polynomial time.  We introduce the concept of a ``decoding light cone''  to establish the local and efficiently measurable nature of this probe.  We also estimate bulk and surface critical exponents for the transition.  Developing scalable probes of MIC in more general models may be a useful application of noisy-intermediate scale quantum (NISQ) devices, as well as point to more efficient realizations of fault-tolerant quantum computation.
\end{abstract}
\maketitle

\emph{Introduction.---}Equilibration to long-time states in many-body systems arises due to entropy production between subsystems and/or with an environment.  
In closed quantum systems that thermalize, this entropy is in the form of long-range entanglement between subsystems \cite{Deutsch91,Srednicki94,Nandkishore15,DAlessio16}.  When a quantum system is coupled to an environment, it is natural to ask whether this entanglement between subsystems can survive coupling to the bath.  If so, this would imply, due to monogamy of entanglement, that there are protected subspaces of the system's state space about which the environment does not gain information during the dynamics \cite{Schumacher96}.  Such a scenario might seem  implausible, however, in some contexts it occurs quite naturally, e.g., in topologically ordered systems \cite{Wen90,Kitaev03} and any realization of a quantum error correcting code \cite{Calderbank96,Steane96,NielsenChuang}.  These basic questions about nonequilibrium quantum statistical mechanics, therefore, have direct relevance to the more practical challenge of realizing fault-tolerant quantum computation \cite{Aharonov00,Gottesman09}.

Recently, it was found that when local unitary entangling dynamics is interspersed with measurements, there is a phase transition between an area-law entangled state in the system at high measurement rate and a volume-law entangled state at low measurement rate \cite{Li18,Skinner18,Chan18b}.  In the area-law phase, equilibration occurs predominantly through entanglement with the local Markovian environment and any long-range entanglement within the system is suppressed, while in the volume-law phase some long-range entanglement between subsystems is produced.   There has  already been significant progress understanding different aspects of this transition, including probes of universal behavior in large classes of models \cite{Li19}, generalizations to weak measurements \cite{Szyniszewski19}, and alternative viewpoints in terms of channel capacities, quantum error correction  \cite{Choi19,Gullans19d}, and purification dynamics \cite{Gullans19d}.  In some limiting cases, the phase transition can be studied analytically in a family of classical statistical mechanical models derived via replica methods \cite{Vasseur18,Bao19,Jian19}.   In these effective models, entanglement is mapped to the free energy cost of inserting a domain wall in the system, raising the question of whether there also local probes that can capture the universal, critical properties of the transition.  Furthermore, the intrinsically random outcomes of quantum measurements prevent one from preparing multiple copies of a single state without either exponentially many samples or potentially complex decoding operations.  As a result, one might suspect that this phase transition is fundamentally inaccessible in experiments with only polynomial resources.

In this Letter, we introduce local, scalable probes of such measurement-induced criticality (MIC) that are immediately applicable to quantum computing platforms with high-fidelity control on large numbers of qubits \cite{Ladd10}.  A central element of our proposal is the identification of a local order parameter for these transitions equal to the entropy of a finite number of maximally entangled reference qubits with the system.  Using this  order parameter, one can extract universal features of the volume-law phase in any spatial dimension and in systems with long-range interactions using constant-depth quantum circuits and polynomially-many runs of the experiment.  Accessing the critical region experimentally requires an efficient method for computing ``entropy decoder functions'' that can correlate the basis of the reference qubits with the measurement record, using an incomplete model for the underlying dynamics of the system.

Using a $1+1$ dimensional stabilizer circuit model that realizes one universality class for MIC \cite{Li19} and can be simulated classically in polynomial time \cite{Gottesman98,Aaronson04}, we show how to identify the critical point with this local order parameter.  We then establish the existence of a decoding light cone defined by the space-time location of measurement events that purify the reference qubits.  We directly show that this local spreading of quantum information into the measurements allows scalable probes of the two phases in large systems.  We then turn to an examination of critical scaling properties of the order parameter.  
As is typical of critical phenomena, the behavior of $n$-point functions in finite-size systems depends sensitively on the underlying  topology \cite{Binder83,Cardy84}.  We illustrate how to use this property to extract a ``surface'' order parameter exponent $\beta_s$.  To measure the ``bulk'' order parameter exponent $\beta$ \cite{Wegner76}, finite size effects are reduced by measuring the two-point function, which we identify with the mutual information between two initially locally entangled reference qubits.  
%All the bulk and surface critical exponents we have extracted for the stabilizer circuit models are close to those of 2D percolation within our estimated uncertainty; however, other surface exponents are known to substantially differ from percolation \cite{Li19,Gullans19d,Li20}.  

\begin{figure}[tb]
\begin{center}
\includegraphics[width = .49 \textwidth]{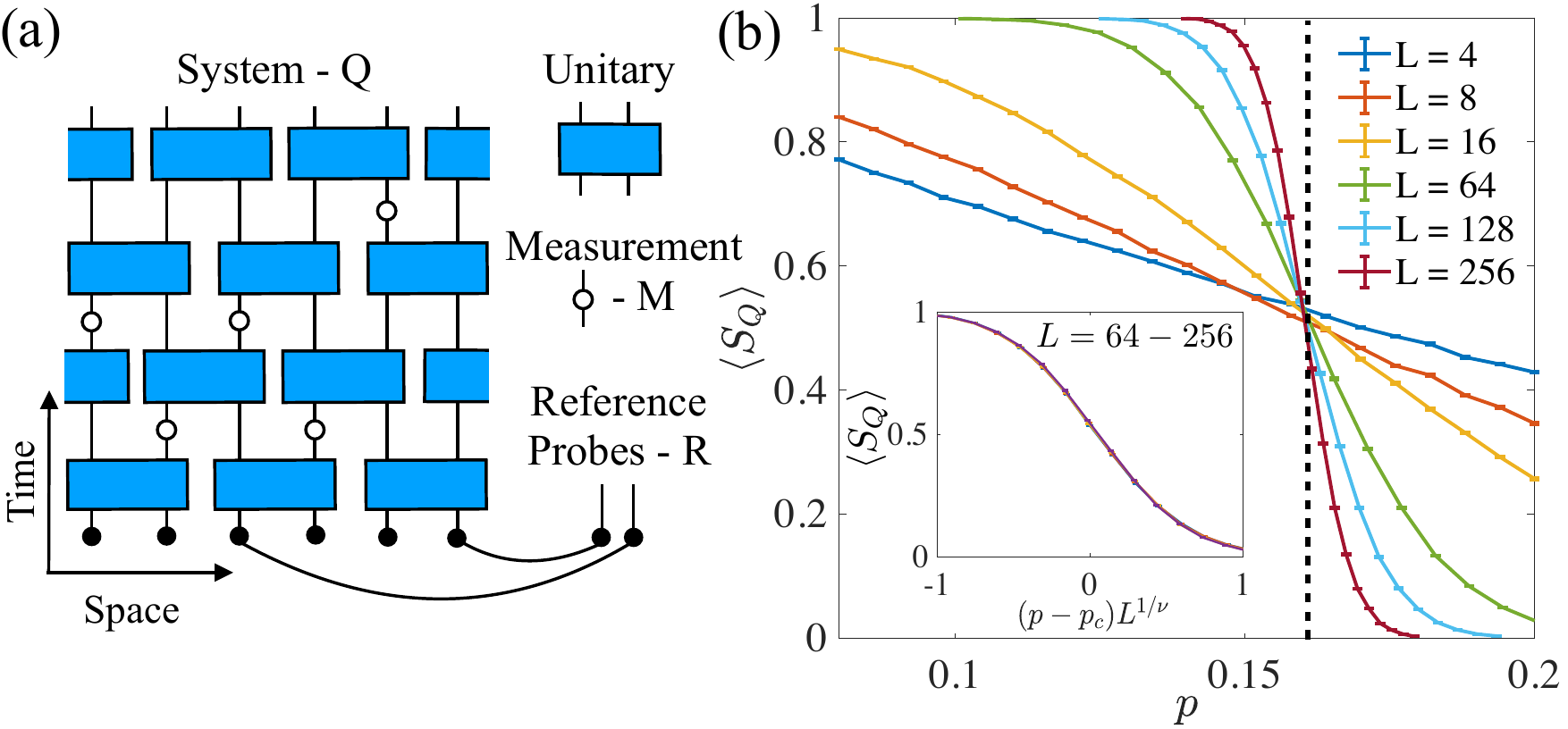}
\caption{(a) Unitary-measurement dynamics in $1+1$ dimensions with additional reference probes.  The reference qubits are used to measure few-point order parameter correlations.  (b) Finite-size scaling of the entanglement transition in a stabilizer circuit model using the circuit-averaged $S_Q$ as an order parameter (see text).  Each two-site unitary is drawn uniformly from the  Clifford group and $Z$-measurements are made at each site with probability $p$.    The crossing point for $L = 64-256$ lets us locate $p_c = 0.1598(5)$ and (inset) a collapse of the data at this value of $p_c$ occurs for $\nu = 1.30(5)$, consistent with previous estimates \cite{Li19,Gullans19d}. }
\label{fig:1}
\end{center}
\end{figure}

\emph{Order parameter measurement.---}Combined unitary-measurement dynamics in one of its simplest form refers to the open system dynamics described by the family of quantum channels 
\begin{align} \label{eqn:Nt}
\mathcal{N}_t(\rho) &= \sum_{\vec{m}} K_{\vec{m}} \rho K_{\vec{m}}^\dag \otimes \ket{\vec{m}} \bra{\vec{m}} , \\
K_{\vec{m}} & = U_t P_t^{m_t} \cdots U_1 P_1^{m_1},
\end{align}
where $\rho$ is the density matrix of the system, $U_n$ are unitary operators, $P_n^{m_n}$ is a sequences of projectors that  satisfy $ P_n^{0}+P_n^1 = \mathbb{I}$,  and $\vec{m}$ indexes the measurement outcomes ($m_n=0$ or 1).  Such channels describe a system that is coupled to the environment only through ancilla qubits, which also act as a register to record the quantum trajectories of the system \cite{Plenio98}.  We note that more general definitions of measurement-induced transitions and phases have been put forward in our recent work \cite{Gullans19d}. We consider an equivalent formulation of the model shown in Fig.~\ref{fig:1}(a),  where the initial density matrix of the system $S$ $\rho_S = \sum_k \lambda_k \ket{k} \bra{k}$ is purified by adding a reference system $R$: $\ket{\psi_{RS}} = \sum_k \sqrt{\lambda_k} \ket{k_R}  \ket{k}$.  In each layer of the circuit, we apply spatially local unitaries, followed by a round of single-site measurements of each site with probability $p$.   For rather generic choices of unitaries, MIC arises in such models by tuning the measurement rate $p$ to a critical value $p_c$.

Previously we showed that one could identify the phase transition by studying the purification dynamics of the maximally mixed state \cite{Gullans19d}; however, the entropy of this mixed state has a similar interpretation to entanglement as a domain wall free-energy cost  \cite{Bao19} and does not serve as a local or scalable probe.  Here, we instead consider the case where the reference system consists of a finite number of qubits.  For simplicity and ease of experimental implementation, we first focus on a single-reference qubit.  
We extend the channel to a unitary operation by including an environment $\mathcal{N}_t(\rho_{S}) = \trace_E[U_{SE} \rho_{SE} U_{SE}^\dag ]$.  The total state of the reference, system,  and environment $\ket{\psi_{RSE}}$ evolves as  
 \begin{align}
\ket{\psi_{RSE}} &=  \sum_{k \vec{m}}  \sqrt{p_{k \vec{m}}} \, \ket{k_R}  \ket{\psi_{k \vec{m}} }   \ket{\vec{m}},
    \end{align}
    where   $\sqrt{ p_{k\vec{m}}} \ket{\psi_{k \vec{m}}} = \sqrt{\lambda_k} (K_{\vec{m}} \ket{k}) \ket{\vec{m}} $ and $p_{k \vec{m}}$ is the  joint probability of starting in $\ket{k}$ and observing measurements $\vec{m}$.  The reduced density matrix for the reference and environment is
$  \rho_{RE}  = \sum_{\vec{m}} p_{\vec{m}}\,  \rho_{R \vec{m}} \otimes\ket{\vec{m}} \bra{\vec{m}}$ with
  \begin{align}
  \rho_{R \vec{m}} & =\left(\begin{array}{c c}
  {p}_{0 |\vec{m}} & \sqrt{ {p}_{0 | \vec{m}} {p}_{1 | \vec{m}}} O_{\vec{m}} \\
  \sqrt{ {p}_{0 | \vec{m}} {p}_{1 |\vec{m}}} O_{\vec{m}}^* & {p}_{1 | \vec{m}}
  \end{array} \right) ,
  \end{align}
  where $p_{\vec{m}}= \sum_{k} p_{k \vec{m}}$, $p_{k| \vec{m}} = p_{k \vec{m}}/ p_{\vec{m}}$ is the conditional probability of the reference being in state $\ket{k_R}$, and $O_{\vec{m}} = \bra{\psi_{0 \vec{m}} }\psi_{1 \vec{m}}\rangle$ is an overlap factor.  We introduce ``quantum'' and ``classical''  order parameters based on this reduced density matrix.  We define the quantum order parameter as the coherent quantum information of this input state \cite{Schumacher96}, which, for the channels in Eq.~(\ref{eqn:Nt}), reduces to the average entropy of the reference qubit  \cite{Gullans19d,Choi19}
 \be
 S_Q =   S(\rho_R) - I(R:E)= \sum_{\vec{m}} p_{\vec{m}} S(\rho_{R \vec{m}}),
 \ee 
 where $S(\rho) = - \trace[\rho \log \rho]$ is the von Neumann entropy and $I(R:E) = S(\rho_R) + S(\rho_E) - S(\rho_{RE})$ is the mutual information.  $S_Q$ measures the ability of the system to store one bit of quantum information  \cite{Schumacher96,HolevoBook}.   
In the ordered phase, the environment gains little information about the state of the reference and $S_Q$ can stay nonzero.  In contrast, in the disordered phase, the environment quickly learns about the state of the reference and $S_Q$ decays to zero.  
 
 To define the classical order parameter $S_C$, we set the off-diagonal elements of $\rho_{R \vec{m}}$ to zero 
  \be
 S_C = H(p_{k \vec{m}}) - H(p_{\vec{m}}) =  \sum_{k \vec{m}} {p}_{k\vec{m}} \log ({p}_{ \vec{m}}/p_{k\vec{ m}}),
 \ee
 where $H(q_i) = - \sum_i q_i \log q_i$ is the classical entropy.     $S_C$ measures the ability of the environment to distinguish the two initial states $\ket{0}$ and $\ket{1}$.  Analogous to $S_Q$, it measures the ability of the system to store one classical bit of information \cite{HolevoBook}.   We remark that a related metric to $S_C$ is the Kullback-Leibler divergence of the measurement distributions for two initial  states $\ket{0}$ and $\ket{1}$
 \be
 D_{\rm KL}(p_{1 \vec{m}} | p_{0\vec{m}}) = \sum_{\vec{m}} p_{1 \vec{m}} \log (p_{1\vec{m}}/p_{0\vec{m}}),
 \ee
 which was identified as a  probe of MIC in Ref.~\cite{Bao19}.  Near the critical point, we expect all of these metrics to have the same universal scaling behavior.
 
   To demonstrate the utility of $\mean{S_Q}$ as a probe of the transition, we turn to the $1+1$ dimensional stabilizer circuit model introduced in Ref.~\cite{Li19}, where each two-site unitary in Fig.~\ref{fig:1}(a) is given by a random Clifford gate and, without loss of generality, each measurement is made along the $Z$ axis.  Stabilizer circuits have the advantage that  efficient classical simulations are straightforward to implement for any dimension or interaction range \cite{Aaronson04}, making them suitable for scalable experiments that include the critical region.

  To identify the critical measurement rate, we initialize systems of length $L$ qubits with periodic boundary conditions by first performing an ``encoding'' step that starts from the reference maximally entangled with one site.  We then create a  pseudo-random stabilizer state by running the circuit without measurements for time $t_0 = 2L$, then run the circuit with measurements for an additional time $t-t_0 = 2L$.
 For $p<p_c$, the entanglement of the system with the reference qubit will be approximately preserved during the dynamics, which leaves $\rho_{R \vec{m}}$ close to a maximally mixed state.  On the other hand, for $p>p_c$, the measurements quickly collapse the entanglement, reducing $\rho_{R \vec{m}}$ to a pure state with either $|O_{\vec m}| \to 1$ or one of ${p}_{k| \vec{m}} \to 0$.  At the critical point, the reference qubit purifies on a timescale $\sim L$ \cite{Gullans19d}. 
 
   In Fig.~\ref{fig:1}(b), we show the finite-size scaling of the circuit-averaged $\mean{S_Q}$ through the entanglement transition.  There is an emergent conformal symmetry in the $1+1$ dimensional models \cite{Li19,Jian19}, which fixes $z = 1$.  We use the scaling ansatz 
    \be \label{eqn:s1}
  \mean{S_Q}  = F[(p-p_c)L^{1/\nu},t/L],
  \ee
  where $t$ is the number of two-qubit gates that have acted on each site.  For this protocol, there is no early time power-law decay because we are quenching the system from the ``ordered'' phase. 
We locate the critical measurement rate $p_c = 0.1598(5)$ through the crossing with increasing system size for $64\le L \le 256$.  Collapsing the data according to Eq.~(\ref{eqn:s1}) with this value of $p_c$ gives an estimate for the correlation length exponent $\nu = 1.30(5)$ \cite{ConInt}.  We find excellent agreement of $p_c$ and $\nu$ with past results \cite{Li19,Gullans19d}.  To illustrate that this approach is applicable to small-scale systems commonly studied in experiments, we include data for $4\le L\le 16$.  With this restricted data set, we obtain similar estimates  $(p_c,\nu) = (0.16(1),1.3(2))$ with less precision.

\emph{Decoding light cone.---}This analysis shows that  we can obtain a direct probe of the phase transition and critical point, provided we can estimate an entropy decoder function:
  \begin{align}
  \vec{m} \to (p_{0| \vec{m}}, O_{\vec{m}}).
  \end{align}
There are three basic approaches to finding this decoder in experiment.  One approach is to implement models, such as stabilizer circuits, that allow efficient classical simulations.  The simulations allow one to make a good guess for the appropriate basis to analyze each measurement result for the reference qubit.  Another approach is to use the experimental data to correlate the measurement record with simultaneous tomography measurements of the reference qubit. This approach allows one to directly reconstruct the decoding function, but could require exponentially many runs of the experiment near the critical point.  A third approach, which we do not explore here, is to use hybrid methods that use the data output from the experiment as input to a classical model for the decoder.  

\begin{figure}[tb]
\begin{center}
\includegraphics[width = 0.49 \textwidth]{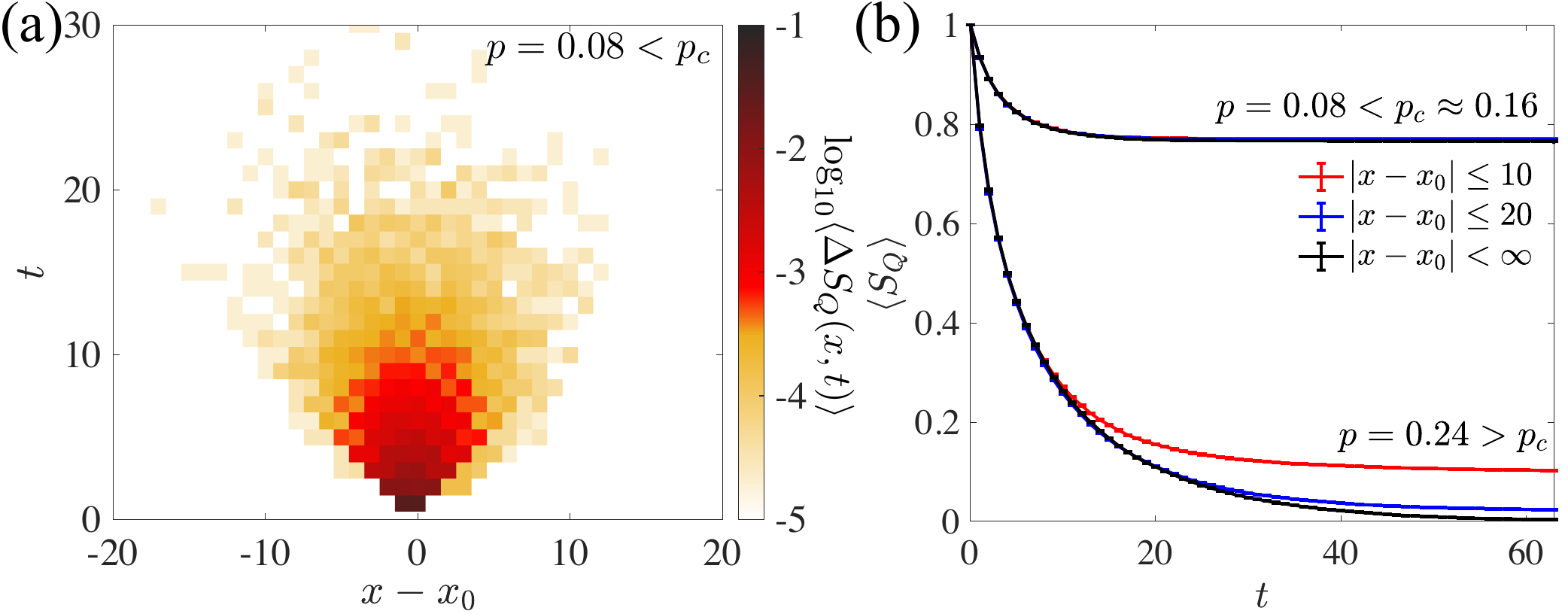}
\caption{(a) Decoding light cone defined by $\mean{\Delta S_Q(x,t)}$, which is the average change in $S_Q$ due to a measurement at space-time point $(x,t)$.  We took periodic boundary conditions with a product initial state and the reference maximally entangled with site $x_0=L/2$ for $L = 64$. (b) Evolution of $\mean{S_Q}$ in the volume and area-law phase for the same condition as (a), but  the measurement results are only recorded when they occur a distance $|x-x_0|$ below the indicated bounds.}
\label{fig:2}
\end{center}
\end{figure}

Although one might suspect that estimating such a decoder is equivalent in difficulty to solving the quantum dynamics of the circuit, this is not generally the case in either of the two phases.  In the volume-law phase, where the overall complexity of the system is highest, the entropy reduction of the reference qubit only takes place on time scales $\sim \xi^z$, where $\xi \sim |p-p_c|^{-\nu}$ is the correlation length of the phase transition and $z$ is the dynamical critical exponent.   After this point, the scrambling dynamics imply that future measurements gain exponentially decreasing amounts of information about the state of the reference.  Thus, we can accurately estimate the decoder in the volume-law phase with a constant-depth quantum circuit.  A second crucial observation is that the decoder only requires access to the measurement record over a bounded space-time domain within the causal lightcone of the reference qubit.   We show an example of this emergent decoding light cone in Fig.~\ref{fig:2}(a) for $p<p_c$ starting from a product initial state with the reference entangled with site $x_0 = L/2$.  
Here, $\mean{\Delta S_Q(x,t)}$ is defined as the average change in $S_Q$ due to a measurement at space-time location $(x,t)$.  
 Perhaps surprisingly, we find the same emergent  light cone for volume-law entangled initial states as long as the reference qubit begins locally entangled with the system \cite{supp}.  In recent work, we introduced a complementary definition of an information spreading light cone in terms of the mutual information of the reference qubit with the system and not the environment \cite{Ippoliti20}.  These locality results further imply that if one reference qubit remains in a mixed state, then an extensive number of them separated by much more than the correlation length will as well.     To further confirm that only a polynomial number of experimental runs are required, we  explicitly model the case where the measurement outcomes are recorded only for $|x-x_0|$ below some cutoff length.  The results  are shown in Fig.~\ref{fig:2}(b) for $p= 0.08  \approx p_c/2$ and $ p = 0.24 \approx 3 p_c/2$.  We find that $S_Q$ converges close to its ideal value as soon as the cutoff exceeds the correlation length. This method explicitly fails at the critical point, where the correlation length diverges; however, in $1+1$ dimensions the entanglement only grows logarithmically in time at $p_c$ \cite{Skinner18}, making decoders based on classical simulation feasible.

%It has been shown that certain limiting cases of this transition are equivalent to 2D critical percolation \cite{Skinner18,Jian19,Bao19}.  Using this mapping as a guide for exploring the critical behavior in the stabilizer circuit models, we note that percolation has a rather small bulk order parameter exponent $\beta = 5/36$ \cite{Percolation}, which makes it difficult to estimate accurately in finite-size systems.  One approach to increase this exponent is to study surface critical behavior.   In the case of percolation, the surface exponent $\beta_s$ measures the probability that a point on the surface is connected to the infinite cluster $P_{sc} \sim |p-p_c|^{\beta_s}$ and takes the much larger value in 2D of $\beta_s=4/9$ \cite{Percolation,Monetti94}.  

\begin{figure}[tb]
\begin{center}
\includegraphics[width = 0.45 \textwidth]{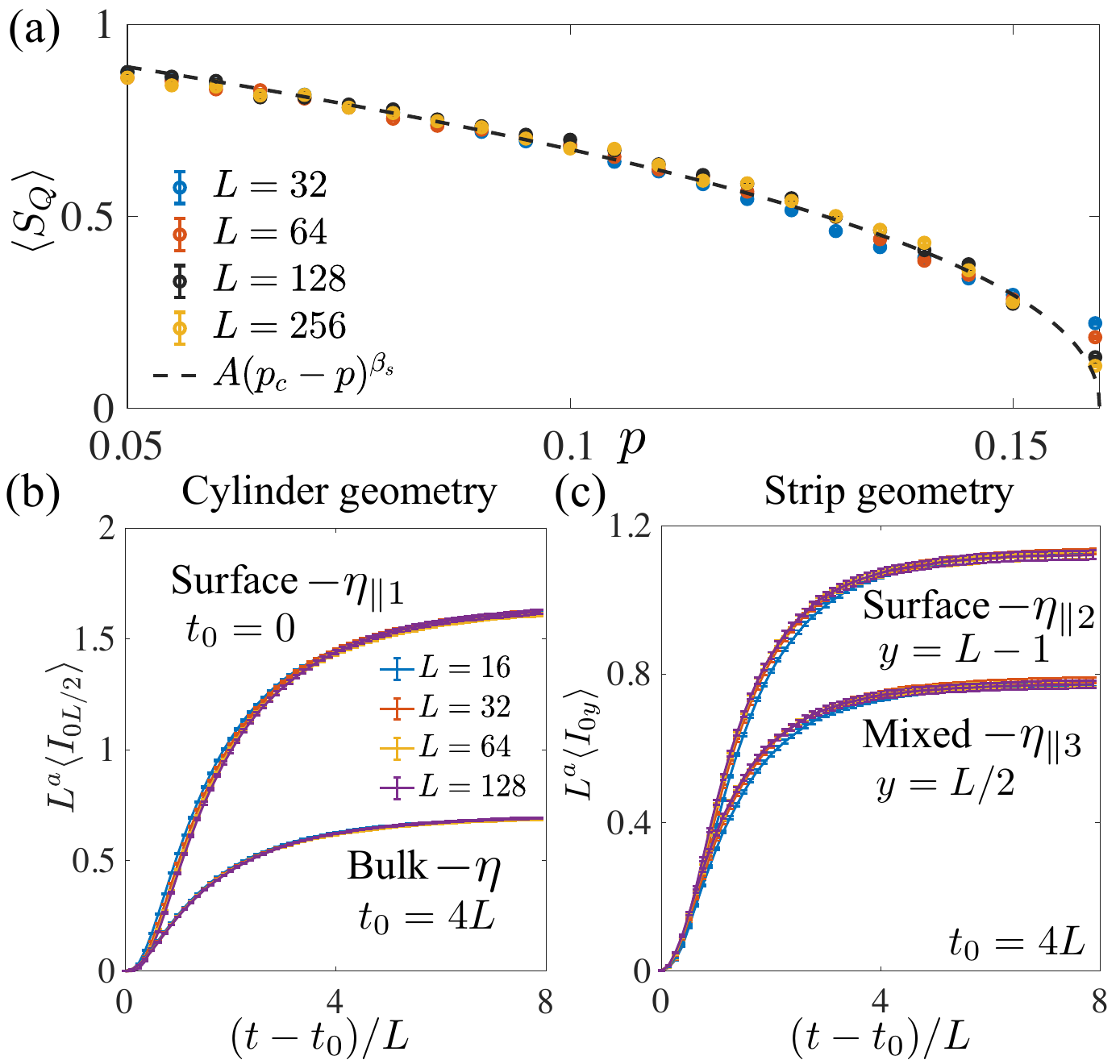}
\caption{(a) $\mean{S_Q}$ when the initial product state has one qubit maximally entangled with the reference, and is then run out to time $t = 2L$.  This procedure allows us to compute the surface order parameter exponent $\beta_s$ from the scaling $\mean{S_Q} \sim |p-p_c|^{\beta_s}$.  (b) Surface/bulk two-point function obtained at $p=0.1596$ by measuring the mutual information $I_{xy}$ between two  reference qubits locally entangled with the system at time $t_0 = 0/4L$ at two antipodal sites $(x,y) = (0,L/2)$ with periodic boundary conditions.  (c) Two-point function for open boundary conditions with $t_0 = 4L$ and $(x,y) = (0,L-1)$ or $(0,L/2)$. We find $(\eta,\eta_{\parallel 1},\eta_{\parallel 2},\eta_{\parallel 3})=(0.22(1),0.74(1),0.67(2), 0.58(2) ).$}
\label{fig:3}
\end{center}
\end{figure}

\emph{Order parameter correlations.---}Having established the possibility of locating the transition with $\mean{S_Q}$, we now turn to the determination of the order parameter critical exponents and  correlation functions.
To use our reference qubit to estimate the surface-order parameter exponent $\beta_s$, we apply a similar procedure as in Fig.~\ref{fig:1}(b), but with the initial state chosen to be a product state and the reference qubit entangled with one of the system's qubits at this ``disordered'' surface.  With this protocol, the reference has a much higher chance of purifying at early times compared to being placed in the bulk.  The numerical results vs.~$p$ are shown in Fig.~\ref{fig:3}(a), where we  compute $\mean{S_Q}$ at time $t = 2L$.  Away from the critical point, we see a collapse of the data with the scaling $\mean{S_Q} \sim |p-p_c|^{\beta_s}$ for $\beta_s  = 0.45(2)$ obtained from fitting.   It has been shown that certain limiting cases of this transition are equivalent to 2D critical percolation \cite{Skinner18,Jian19,Bao19}. Interestingly, our extracted value of $\beta_s$ is  close to the percolation value $\beta_s = 4/9$ \cite{Percolation,Monetti94}. 

% In Fig.~\ref{fig:2}(b), we perform a similar analysis at the critical point by looking at the scaling behavior of $\mean{S_Q}$ with time.  At intermediate times $1 \ll t \lesssim L$, it decays with a  power-law $\mean{S_Q} \sim t^{-\eta_\parallel/2}$ before crossing over to a universal decay [see inset to  Fig.~\ref{fig:2}(b)] as a function of $t/L$ for times $t \gtrsim L$.  The exponent $\eta_\parallel$ satisfies the scaling relation $\eta_\parallel = 2 \beta_s/\nu$ \cite{Binder83,Monetti94}.  From fitting the power law decay at intermediate times, we extract $\eta_\parallel = 0.67(2)$, consistent with the value computed from $\beta_s$ and $\nu$.

We can obtain an accurate probe of the bulk order parameter exponent by measuring connected two-point order parameter correlation functions using an additional reference qubit.
At time $t_0$, we measure two qubits in the system at positions $x$ and $y$ and then place each one of these qubits in a maximally entangled state with a reference qubit.  We then compute the mutual information between the two reference qubits $I_{xy}$ as a function of $(t-t_0)/L$.  Scaling theory predicts that the circuit averaged $\mean{I_{xy}}$ at $p=p_c$ should have the form \cite{Wegner76}
\be
\mean{I_{xy}} = |x-y|^{-\eta} G[ (t-t_0)/L],
\ee
for a universal scaling function $G(\cdot)$.  In Fig.~\ref{fig:2}(b), we show that $\mean{I_{xy}}$ follows precisely this predicted form when  $t_0 = 0/4L$ and  $x$ and $y$ are taken to be antipodal sites ($|x-y| = L/2$) in a system with periodic boundary conditions.  Here, we are essentially measuring two-point functions in the surface/bulk of a cylinder.  From collapsing the data, we obtain a bulk exponent $\eta  = 0.22(1)$ (close to percolation $\eta = 5/24$) and surface exponent $\eta_{\parallel 1} = 0.74(1)$ (as compared to percolation  $\eta_{\parallel} = 2/3$).   Using open boundary conditions, we can obtain independent estimates of the surface exponent as shown in Fig.~\ref{fig:2}(c).  Placing one reference qubit at $x = 0$ and another at $y = L-1$,  we obtain the estimate  $\eta_{\parallel 2} = 0.67(2)$ \cite{conf}.  For $y = L/2$,  scaling theory predicts an exponent $\eta_{\perp} = (\eta+\eta_{\parallel})/2$ \cite{Binder83,Cardy84}, from which we estimate a third value $\eta_{\parallel 3} = 0.58(2)$.  Using these differences to estimate systematic uncertainty, we obtain the estimate $\eta_{\parallel} = 0.7(1)$.
We note that, recently, Ref.~\cite{Li20} studied this stabilizer circuit model and found a slightly larger surface exponent $(\eta_{\parallel }=0.82)$ using the purification dynamics of several reference qubits (see supplemental material for a discussion of this estimate \cite{supp}).  Moreover, other surface exponents in this stabilizer circuit  model are known to have much more substantial differences from percolation \cite{Li19,Gullans19d,Li20}.  Finally, we note that recent work on a 1+1 dimensional  Haar random circuit with measurements found a similar value of $\eta = 0.19(1)$, but a substantially different value $\eta_{\parallel 2} = 0.39(1)$ \cite{Zabalo20}. 

% Working with a single planar surface introduces two new critical exponents for order parameter correlations parallel $(\eta_{\parallel})$ and perpendicular $(\eta_\perp)$ to the surface \cite{Binder83}.  We measure $\eta_\parallel$ by starting from a product state and setting $\tau = 0$.  The results are shown in the inset to Fig.~\ref{fig:2}(c).  There are strong finite-size effects, but, at the large sizes, we find consistent scaling results with $\eta_\parallel$ extracted from Fig.~\ref{fig:2}(b).  We have estimated (data not shown) $\eta_\perp = 0.43(2)$ by performing a similar analysis with open boundary conditions, $(x,y) = (0,L/2)$, and $\tau = 4 L$.  Our measured exponents are consistent with the   predicted scaling relation $2 \eta_\perp = \eta + \eta_\parallel$ \cite{Binder83,Cardy84}.

\emph{Conclusions.---}We have defined a local order parameter for MIC and shown how it can be used to realize scalable probes of this novel class of critical phenomena.  Our proposals are immediately applicable to quantum computing platforms with high-fidelity control on large numbers of qubits.  Although we focused on a $1+1$ dimensional stabilizer circuit model, the proposed methodology can be applied to any known realization of MIC in any number of dimensions or range of interactions.  In cases with long-range interactions, entanglement within the system may no longer be a useful diagnostic of the phase transition, but MIC is still realized in the purification dynamics of the reference system \cite{Gullans19d}.  

Many open questions remain about the appropriate classification of these phase transitions, especially outside $1+1$ dimensions or in the presence of quenched disorder.  The ordered phase naturally realizes high complexity states, which raises questions about the relation of MIC to quantum complexity theory.  As a result, developing scalable probes of MIC in more general models may be a useful application of noisy-intermediate scale quantum  (NISQ) devices \cite{Preskill18}.  We have found that our order parameter can be extracted from the entropy of measurement outcomes in a fixed basis, which can be directly estimated using techniques similar to  cross-entropy benchmarking \cite{Boixo18}. Furthermore, the ordered phase naturally realizes novel quantum error correcting codes \cite{Choi19,Gullans19d,Fan20}.  Studying the properties of these codes, including their universal scaling properties near the transition, may provide fundamental insights into quantum error correction,  potentially pointing to more efficient realizations of fault-tolerant quantum computation.  

 \begin{acknowledgments}
We thank  Ehud Altman, Soonwon Choi,  Steve Flammia, Steve Girvin, Sarang Gopalakrishnan, Alexey Gorshkov, Matteo Ippoliti, Liang Jiang, Vedika Khemani, Stefan Krastanov,  Pradeep Niroula, Crystal Noel, Chris Monroe,  Jed Pixley,  Romain Vasseur, Justin Wilson, and Aidan Zabalo for helpful discussions.  Research supported in part by the DARPA DRINQS program.  D.A.H. was also supported in part by a Simons Fellowship.
%Research supported in part by
%DARPA grant No. D18AC0025,  and the Gordon and Betty Moore Foundation's EPiQS Initiative through Grant GBMF4535.  
\end{acknowledgments}

\bibliographystyle{apsrev-nourl-title-PRX}
\bibliography{LocalProbes}

\begin{thebibliography}{41}
\expandafter\ifx\csname natexlab\endcsname\relax\def\natexlab#1{#1}\fi
\expandafter\ifx\csname bibnamefont\endcsname\relax
  \def\bibnamefont#1{#1}\fi
\expandafter\ifx\csname bibfnamefont\endcsname\relax
  \def\bibfnamefont#1{#1}\fi
\expandafter\ifx\csname citenamefont\endcsname\relax
  \def\citenamefont#1{#1}\fi
\expandafter\ifx\csname url\endcsname\relax
  \def\url#1{\texttt{#1}}\fi
\expandafter\ifx\csname urlprefix\endcsname\relax\def\urlprefix{URL }\fi
\providecommand{\bibinfo}[2]{#2}
\providecommand{\eprint}[2][]{\url{#2}}

\bibitem[{\citenamefont{Deutsch}(1991)}]{Deutsch91}
\bibinfo{author}{\bibfnamefont{J.~M.} \bibnamefont{Deutsch}},
  \emph{\bibinfo{title}{Quantum statistical mechanics in a closed system}},
  \bibinfo{journal}{Phys. Rev. A} \textbf{\bibinfo{volume}{43}},
  \bibinfo{pages}{2046} (\bibinfo{year}{1991}).

\bibitem[{\citenamefont{Srednicki}(1994)}]{Srednicki94}
\bibinfo{author}{\bibfnamefont{M.}~\bibnamefont{Srednicki}},
  \emph{\bibinfo{title}{Chaos and quantum thermalization}},
  \bibinfo{journal}{Phys. Rev. E} \textbf{\bibinfo{volume}{50}},
  \bibinfo{pages}{888} (\bibinfo{year}{1994}).

\bibitem[{\citenamefont{Nandkishore and Huse}(2015)}]{Nandkishore15}
\bibinfo{author}{\bibfnamefont{R.}~\bibnamefont{Nandkishore}} \bibnamefont{and}
  \bibinfo{author}{\bibfnamefont{D.~A.} \bibnamefont{Huse}},
  \emph{\bibinfo{title}{{Many-body localization and thermalization in quantum
  statistical mechanics}}}, \bibinfo{journal}{Annu. Rev. Condens. Matter Phys.}
  \textbf{\bibinfo{volume}{6}}, \bibinfo{pages}{15} (\bibinfo{year}{2015}).

\bibitem[{\citenamefont{D'Alessio et~al.}(2016)\citenamefont{D'Alessio, Kafri,
  Polkovnikov, and Rigol}}]{DAlessio16}
\bibinfo{author}{\bibfnamefont{L.}~\bibnamefont{D'Alessio}},
  \bibinfo{author}{\bibfnamefont{Y.}~\bibnamefont{Kafri}},
  \bibinfo{author}{\bibfnamefont{A.}~\bibnamefont{Polkovnikov}},
  \bibnamefont{and} \bibinfo{author}{\bibfnamefont{M.}~\bibnamefont{Rigol}},
  \emph{\bibinfo{title}{{From quantum chaos and eigenstate thermalization to
  statistical mechanics and thermodynamics}}}, \bibinfo{journal}{Adv. Phys.}
  \textbf{\bibinfo{volume}{65}}, \bibinfo{pages}{239} (\bibinfo{year}{2016}).

\bibitem[{\citenamefont{Schumacher and Nielsen}(1996)}]{Schumacher96}
\bibinfo{author}{\bibfnamefont{B.}~\bibnamefont{Schumacher}} \bibnamefont{and}
  \bibinfo{author}{\bibfnamefont{M.~A.} \bibnamefont{Nielsen}},
  \emph{\bibinfo{title}{{Quantum data processing and error correction}}},
  \bibinfo{journal}{Phys. Rev. A} \textbf{\bibinfo{volume}{54}},
  \bibinfo{pages}{2629} (\bibinfo{year}{1996}).

\bibitem[{\citenamefont{Wen and Niu}(1990)}]{Wen90}
\bibinfo{author}{\bibfnamefont{X.~G.} \bibnamefont{Wen}} \bibnamefont{and}
  \bibinfo{author}{\bibfnamefont{Q.}~\bibnamefont{Niu}},
  \emph{\bibinfo{title}{Ground-state degeneracy of the fractional quantum hall
  states in the presence of a random potential and on high-genus riemann
  surfaces}}, \bibinfo{journal}{Phys. Rev. B} \textbf{\bibinfo{volume}{41}},
  \bibinfo{pages}{9377} (\bibinfo{year}{1990}).

\bibitem[{\citenamefont{Kitaev}(2003)}]{Kitaev03}
\bibinfo{author}{\bibfnamefont{A.~Y.} \bibnamefont{Kitaev}},
  \emph{\bibinfo{title}{{Fault-tolerant quantum computation by anyons}}},
  \bibinfo{journal}{Ann. Phys.} \textbf{\bibinfo{volume}{303}},
  \bibinfo{pages}{2} (\bibinfo{year}{2003}).

\bibitem[{\citenamefont{Calderbank and Shor}(1996)}]{Calderbank96}
\bibinfo{author}{\bibfnamefont{A.~R.} \bibnamefont{Calderbank}}
  \bibnamefont{and} \bibinfo{author}{\bibfnamefont{P.~W.} \bibnamefont{Shor}},
  \emph{\bibinfo{title}{Good quantum error-correcting codes exist}},
  \bibinfo{journal}{Phys. Rev. A} \textbf{\bibinfo{volume}{54}},
  \bibinfo{pages}{1098} (\bibinfo{year}{1996}).

\bibitem[{\citenamefont{Steane}(1996)}]{Steane96}
\bibinfo{author}{\bibfnamefont{A.~M.} \bibnamefont{Steane}},
  \emph{\bibinfo{title}{{Multiple-particle interference and quantum error
  correction}}}, \bibinfo{journal}{Proc. R. Soc. Lond. A}
  \textbf{\bibinfo{volume}{452}}, \bibinfo{pages}{2551} (\bibinfo{year}{1996}).

\bibitem[{\citenamefont{Nielsen and Chuang}(2011)}]{NielsenChuang}
\bibinfo{author}{\bibfnamefont{M.~A.} \bibnamefont{Nielsen}} \bibnamefont{and}
  \bibinfo{author}{\bibfnamefont{I.~L.} \bibnamefont{Chuang}},
  \emph{\bibinfo{title}{Quantum Computation and Quantum Information}}
  (\bibinfo{publisher}{Cambridge University Press}, \bibinfo{address}{New York,
  NY, USA}, \bibinfo{year}{2011}), \bibinfo{edition}{10th} ed.

\bibitem[{\citenamefont{Aharonov}(2000)}]{Aharonov00}
\bibinfo{author}{\bibfnamefont{D.}~\bibnamefont{Aharonov}},
  \emph{\bibinfo{title}{{Quantum to classical phase transition in noisy quantum
  computers}}}, \bibinfo{journal}{Phys. Rev. A} \textbf{\bibinfo{volume}{62}},
  \bibinfo{pages}{062311} (\bibinfo{year}{2000}).

\bibitem[{\citenamefont{Gottesman}(2009)}]{Gottesman09}
\bibinfo{author}{\bibfnamefont{D.}~\bibnamefont{Gottesman}},
  \emph{\bibinfo{title}{{An Introduction to Quantum Error Correction and
  Fault-Tolerant Quantum Computation}}} (\bibinfo{year}{2009}),
  \eprint{arXiv:0904.2557}.

\bibitem[{\citenamefont{Li et~al.}(2018)\citenamefont{Li, Chen, and
  Fisher}}]{Li18}
\bibinfo{author}{\bibfnamefont{Y.}~\bibnamefont{Li}},
  \bibinfo{author}{\bibfnamefont{X.}~\bibnamefont{Chen}}, \bibnamefont{and}
  \bibinfo{author}{\bibfnamefont{M.~P.~A.} \bibnamefont{Fisher}},
  \emph{\bibinfo{title}{Quantum zeno effect and the many-body entanglement
  transition}}, \bibinfo{journal}{Phys. Rev. B} \textbf{\bibinfo{volume}{98}},
  \bibinfo{pages}{205136} (\bibinfo{year}{2018}).

\bibitem[{\citenamefont{Skinner et~al.}(2019)\citenamefont{Skinner, Ruhman, and
  Nahum}}]{Skinner18}
\bibinfo{author}{\bibfnamefont{B.}~\bibnamefont{Skinner}},
  \bibinfo{author}{\bibfnamefont{J.}~\bibnamefont{Ruhman}}, \bibnamefont{and}
  \bibinfo{author}{\bibfnamefont{A.}~\bibnamefont{Nahum}},
  \emph{\bibinfo{title}{Measurement-induced phase transitions in the dynamics
  of entanglement}}, \bibinfo{journal}{Phys. Rev. X}
  \textbf{\bibinfo{volume}{9}}, \bibinfo{pages}{031009} (\bibinfo{year}{2019}).

\bibitem[{\citenamefont{Chan et~al.}(2019)\citenamefont{Chan, Nandkishore,
  Pretko, and Smith}}]{Chan18b}
\bibinfo{author}{\bibfnamefont{A.}~\bibnamefont{Chan}},
  \bibinfo{author}{\bibfnamefont{R.~M.} \bibnamefont{Nandkishore}},
  \bibinfo{author}{\bibfnamefont{M.}~\bibnamefont{Pretko}}, \bibnamefont{and}
  \bibinfo{author}{\bibfnamefont{G.}~\bibnamefont{Smith}},
  \emph{\bibinfo{title}{Unitary-projective entanglement dynamics}},
  \bibinfo{journal}{Phys. Rev. B} \textbf{\bibinfo{volume}{99}},
  \bibinfo{pages}{224307} (\bibinfo{year}{2019}).

\bibitem[{\citenamefont{Li et~al.}(2019)\citenamefont{Li, Chen, and
  Fisher}}]{Li19}
\bibinfo{author}{\bibfnamefont{Y.}~\bibnamefont{Li}},
  \bibinfo{author}{\bibfnamefont{X.}~\bibnamefont{Chen}}, \bibnamefont{and}
  \bibinfo{author}{\bibfnamefont{M.~P.~A.} \bibnamefont{Fisher}},
  \emph{\bibinfo{title}{Measurement-driven entanglement transition in hybrid
  quantum circuits}}, \bibinfo{journal}{Phys. Rev. B}
  \textbf{\bibinfo{volume}{100}}, \bibinfo{pages}{134306}
  (\bibinfo{year}{2019}).

\bibitem[{\citenamefont{Szyniszewski et~al.}(2019)\citenamefont{Szyniszewski,
  Romito, and Schomerus}}]{Szyniszewski19}
\bibinfo{author}{\bibfnamefont{M.}~\bibnamefont{Szyniszewski}},
  \bibinfo{author}{\bibfnamefont{A.}~\bibnamefont{Romito}}, \bibnamefont{and}
  \bibinfo{author}{\bibfnamefont{H.}~\bibnamefont{Schomerus}},
  \emph{\bibinfo{title}{Entanglement transition from variable-strength weak
  measurements}}, \bibinfo{journal}{Phys. Rev. B}
  \textbf{\bibinfo{volume}{100}}, \bibinfo{pages}{064204}
  (\bibinfo{year}{2019}).

\bibitem[{\citenamefont{Choi et~al.}(2019)\citenamefont{Choi, Bao, Qi, and
  Altman}}]{Choi19}
\bibinfo{author}{\bibfnamefont{S.}~\bibnamefont{Choi}},
  \bibinfo{author}{\bibfnamefont{Y.}~\bibnamefont{Bao}},
  \bibinfo{author}{\bibfnamefont{X.-L.} \bibnamefont{Qi}}, \bibnamefont{and}
  \bibinfo{author}{\bibfnamefont{E.}~\bibnamefont{Altman}},
  \emph{\bibinfo{title}{{Quantum error correction and entanglement phase
  transition in random unitary circuits with projective measurements}}}
  (\bibinfo{year}{2019}), \eprint{arXiv:1903.05124}.

\bibitem[{\citenamefont{Gullans and Huse}(2019)}]{Gullans19d}
\bibinfo{author}{\bibfnamefont{M.~J.} \bibnamefont{Gullans}} \bibnamefont{and}
  \bibinfo{author}{\bibfnamefont{D.~A.} \bibnamefont{Huse}},
  \emph{\bibinfo{title}{{Dynamical purification phase transition induced by
  quantum measurements}}} (\bibinfo{year}{2019}), \eprint{arXiv:1905.05195}.

\bibitem[{\citenamefont{Vasseur et~al.}(2018)\citenamefont{Vasseur, Potter,
  You, and Ludwig}}]{Vasseur18}
\bibinfo{author}{\bibfnamefont{R.}~\bibnamefont{Vasseur}},
  \bibinfo{author}{\bibfnamefont{A.~C.} \bibnamefont{Potter}},
  \bibinfo{author}{\bibfnamefont{Y.-Z.} \bibnamefont{You}}, \bibnamefont{and}
  \bibinfo{author}{\bibfnamefont{A.~W.~W.} \bibnamefont{Ludwig}},
  \emph{\bibinfo{title}{{Entanglement Transitions from Holographic Random
  Tensor Networks}}} (\bibinfo{year}{2018}), \eprint{arXiv:1807.07082}.

\bibitem[{\citenamefont{Bao et~al.}(2019)\citenamefont{Bao, Choi, and
  Altman}}]{Bao19}
\bibinfo{author}{\bibfnamefont{Y.}~\bibnamefont{Bao}},
  \bibinfo{author}{\bibfnamefont{S.}~\bibnamefont{Choi}}, \bibnamefont{and}
  \bibinfo{author}{\bibfnamefont{E.}~\bibnamefont{Altman}},
  \emph{\bibinfo{title}{{Theory of the Phase Transition in Random Unitary
  Circuits with Measurements}}} (\bibinfo{year}{2019}),
  \eprint{arXiv:1908.04305}.

\bibitem[{\citenamefont{Jian et~al.}(2019)\citenamefont{Jian, You, Vasseur, and
  Ludwig}}]{Jian19}
\bibinfo{author}{\bibfnamefont{C.-M.} \bibnamefont{Jian}},
  \bibinfo{author}{\bibfnamefont{Y.-Z.} \bibnamefont{You}},
  \bibinfo{author}{\bibfnamefont{R.}~\bibnamefont{Vasseur}}, \bibnamefont{and}
  \bibinfo{author}{\bibfnamefont{A.~W.~W.} \bibnamefont{Ludwig}},
  \emph{\bibinfo{title}{{Measurement-induced criticality in random quantum
  circuits}}} (\bibinfo{year}{2019}), \eprint{arXiv:1908.08051}.

\bibitem[{\citenamefont{Ladd et~al.}(2010)\citenamefont{Ladd, Jelezko,
  Laflamme, Nakamura, Monroe, and O'Brien}}]{Ladd10}
\bibinfo{author}{\bibfnamefont{T.~D.} \bibnamefont{Ladd}},
  \bibinfo{author}{\bibfnamefont{F.}~\bibnamefont{Jelezko}},
  \bibinfo{author}{\bibfnamefont{R.}~\bibnamefont{Laflamme}},
  \bibinfo{author}{\bibfnamefont{Y.}~\bibnamefont{Nakamura}},
  \bibinfo{author}{\bibfnamefont{C.}~\bibnamefont{Monroe}}, \bibnamefont{and}
  \bibinfo{author}{\bibfnamefont{J.~L.} \bibnamefont{O'Brien}},
  \emph{\bibinfo{title}{{Quantum computers}}}, \bibinfo{journal}{Nature}
  \textbf{\bibinfo{volume}{464}}, \bibinfo{pages}{45} (\bibinfo{year}{2010}).

\bibitem[{\citenamefont{Gottesman}(1998)}]{Gottesman98}
\bibinfo{author}{\bibfnamefont{D.}~\bibnamefont{Gottesman}},
  \emph{\bibinfo{title}{{The Heisenberg Representation of Quantum Computers}}}
  (\bibinfo{year}{1998}), \eprint{arXiv:quant-ph/9807006}.

\bibitem[{\citenamefont{Aaronson and Gottesman}(2004)}]{Aaronson04}
\bibinfo{author}{\bibfnamefont{S.}~\bibnamefont{Aaronson}} \bibnamefont{and}
  \bibinfo{author}{\bibfnamefont{D.}~\bibnamefont{Gottesman}},
  \emph{\bibinfo{title}{Improved simulation of stabilizer circuits}},
  \bibinfo{journal}{Phys. Rev. A} \textbf{\bibinfo{volume}{70}},
  \bibinfo{pages}{052328} (\bibinfo{year}{2004}).

\bibitem[{Bin()}]{Binder83}
\bibinfo{note}{K. Binder, \textit{Phase Transitions and Critical Phenomena}
  vol. 8, ed. C. Domb and J. L. Lebowitz (New York: Academic) (1983).}

\bibitem[{\citenamefont{Cardy}(1984)}]{Cardy84}
\bibinfo{author}{\bibfnamefont{J.~L.} \bibnamefont{Cardy}},
  \emph{\bibinfo{title}{{Conformal invariance and surface critical behavior}}},
  \bibinfo{journal}{Nucl. Phys. B} \textbf{\bibinfo{volume}{240}},
  \bibinfo{pages}{514} (\bibinfo{year}{1984}).

\bibitem[{Weg()}]{Wegner76}
\bibinfo{note}{F. J. Wegner, \textit{Phase Transitions and Critical Phenomena}
  vol. 6, ed. C. Domb and M. S. Green (New York: Academic) (1976).}

\bibitem[{\citenamefont{Plenio and Knight}(1998)}]{Plenio98}
\bibinfo{author}{\bibfnamefont{M.~B.} \bibnamefont{Plenio}} \bibnamefont{and}
  \bibinfo{author}{\bibfnamefont{P.~L.} \bibnamefont{Knight}},
  \emph{\bibinfo{title}{The quantum-jump approach to dissipative dynamics in
  quantum optics}}, \bibinfo{journal}{Rev. Mod. Phys.}
  \textbf{\bibinfo{volume}{70}}, \bibinfo{pages}{101} (\bibinfo{year}{1998}).

\bibitem[{\citenamefont{Holevo}(2012)}]{HolevoBook}
\bibinfo{author}{\bibfnamefont{A.~S.} \bibnamefont{Holevo}},
  \emph{\bibinfo{title}{Quantum Systems, Channels, Information}}
  (\bibinfo{publisher}{Walter de Gruyter GmbH, Berlin/Boston},
  \bibinfo{year}{2012}).

\bibitem[{Con()}]{ConInt}
\bibinfo{note}{Numbers in parentheses denote estimated confidence intervals for
  critical parameters arising from systematic errors in scaling collapses.}

\bibitem[{sup()}]{supp}
\bibinfo{note}{See Supplementary Material for a discussion of the decoding
  light cone with volume-law entangled initial states and analysis of the
  purification dynamics of several reference qubits.}

\bibitem[{\citenamefont{Ippoliti et~al.}(2020)\citenamefont{Ippoliti, Gullans,
  Gopalakrishnan, Huse, and Khemani}}]{Ippoliti20}
\bibinfo{author}{\bibfnamefont{M.}~\bibnamefont{Ippoliti}},
  \bibinfo{author}{\bibfnamefont{M.~J.} \bibnamefont{Gullans}},
  \bibinfo{author}{\bibfnamefont{S.}~\bibnamefont{Gopalakrishnan}},
  \bibinfo{author}{\bibfnamefont{D.~A.} \bibnamefont{Huse}}, \bibnamefont{and}
  \bibinfo{author}{\bibfnamefont{V.}~\bibnamefont{Khemani}},
  \emph{\bibinfo{title}{{Entanglement phase transitions in measurement-only
  dynamics}}} (\bibinfo{year}{2020}), \eprint{arXiv:2004.09560}.

\bibitem[{\citenamefont{Stauffer and Aharony}(2003)}]{Percolation}
\bibinfo{author}{\bibfnamefont{D.}~\bibnamefont{Stauffer}} \bibnamefont{and}
  \bibinfo{author}{\bibfnamefont{A.}~\bibnamefont{Aharony}},
  \emph{\bibinfo{title}{Introduction to Percolation Theory}}
  (\bibinfo{publisher}{Taylor and Francis, London/Philadelphia},
  \bibinfo{year}{2003}).

\bibitem[{\citenamefont{Monetti and Albano}(1994)}]{Monetti94}
\bibinfo{author}{\bibfnamefont{R.~A.} \bibnamefont{Monetti}} \bibnamefont{and}
  \bibinfo{author}{\bibfnamefont{E.~V.} \bibnamefont{Albano}},
  \emph{\bibinfo{title}{{Multiscaling behavior in the crossover between surface
  and bulk critical exponents for percolation in two dimensions}}},
  \bibinfo{journal}{Phys. Rev. E} \textbf{\bibinfo{volume}{49}},
  \bibinfo{pages}{199} (\bibinfo{year}{1994}).

\bibitem[{con()}]{conf}
\bibinfo{note}{In the scaling limit, the geometry for $\eta_{\parallel 2}$ is
  equivalent to the geometry for $\eta_{\parallel 1}$ through a conformal
  transformation of an infinite strip to the upper half plane (see also
  Ref.~\cite{Zabalo20}).}

\bibitem[{\citenamefont{Li et~al.}(2020)\citenamefont{Li, Chen, Ludwig, and
  Fisher}}]{Li20}
\bibinfo{author}{\bibfnamefont{Y.}~\bibnamefont{Li}},
  \bibinfo{author}{\bibfnamefont{X.}~\bibnamefont{Chen}},
  \bibinfo{author}{\bibfnamefont{A.~W.~W.} \bibnamefont{Ludwig}},
  \bibnamefont{and} \bibinfo{author}{\bibfnamefont{M.~P.~A.}
  \bibnamefont{Fisher}}, \emph{\bibinfo{title}{{Conformal invariance and
  quantum non-locality in hybrid quantum circuits}}} (\bibinfo{year}{2020}),
  \eprint{arXiv:2003.12721}.

\bibitem[{\citenamefont{Zabalo et~al.}(2020)\citenamefont{Zabalo, Gullans,
  Wilson, Gopalakrishnan, Huse, and Pixley}}]{Zabalo20}
\bibinfo{author}{\bibfnamefont{A.}~\bibnamefont{Zabalo}},
  \bibinfo{author}{\bibfnamefont{M.~J.} \bibnamefont{Gullans}},
  \bibinfo{author}{\bibfnamefont{J.~H.} \bibnamefont{Wilson}},
  \bibinfo{author}{\bibfnamefont{S.}~\bibnamefont{Gopalakrishnan}},
  \bibinfo{author}{\bibfnamefont{D.~A.} \bibnamefont{Huse}}, \bibnamefont{and}
  \bibinfo{author}{\bibfnamefont{J.~H.} \bibnamefont{Pixley}},
  \emph{\bibinfo{title}{Critical properties of the measurement-induced
  transition in random quantum circuits}}, \bibinfo{journal}{Phys. Rev. B}
  \textbf{\bibinfo{volume}{101}}, \bibinfo{pages}{060301}
  (\bibinfo{year}{2020}).

\bibitem[{\citenamefont{Preskill}(2018)}]{Preskill18}
\bibinfo{author}{\bibfnamefont{J.}~\bibnamefont{Preskill}},
  \emph{\bibinfo{title}{Quantum {C}omputing in the {NISQ} era and beyond}},
  \bibinfo{journal}{{Quantum}} \textbf{\bibinfo{volume}{2}},
  \bibinfo{pages}{79} (\bibinfo{year}{2018}).

\bibitem[{\citenamefont{Boixo et~al.}(2018)\citenamefont{Boixo, Isakov,
  Smelyanskiy, Babbush, Ding, Jiang, Bremner, Martinis, and Neven}}]{Boixo18}
\bibinfo{author}{\bibfnamefont{S.}~\bibnamefont{Boixo}},
  \bibinfo{author}{\bibfnamefont{S.~V.} \bibnamefont{Isakov}},
  \bibinfo{author}{\bibfnamefont{V.~N.} \bibnamefont{Smelyanskiy}},
  \bibinfo{author}{\bibfnamefont{R.}~\bibnamefont{Babbush}},
  \bibinfo{author}{\bibfnamefont{N.}~\bibnamefont{Ding}},
  \bibinfo{author}{\bibfnamefont{Z.}~\bibnamefont{Jiang}},
  \bibinfo{author}{\bibfnamefont{M.~J.} \bibnamefont{Bremner}},
  \bibinfo{author}{\bibfnamefont{J.~M.} \bibnamefont{Martinis}},
  \bibnamefont{and} \bibinfo{author}{\bibfnamefont{H.}~\bibnamefont{Neven}},
  \emph{\bibinfo{title}{{Characterizing quantum supremacy in near-term
  devices}}}, \bibinfo{journal}{Nature Phys.} \textbf{\bibinfo{volume}{14}},
  \bibinfo{pages}{595} (\bibinfo{year}{2018}).

\bibitem[{\citenamefont{Fan et~al.}(2020)\citenamefont{Fan, Vijay, Vishwanath,
  and You}}]{Fan20}
\bibinfo{author}{\bibfnamefont{R.}~\bibnamefont{Fan}},
  \bibinfo{author}{\bibfnamefont{S.}~\bibnamefont{Vijay}},
  \bibinfo{author}{\bibfnamefont{A.}~\bibnamefont{Vishwanath}},
  \bibnamefont{and} \bibinfo{author}{\bibfnamefont{Y.-Z.} \bibnamefont{You}},
  \emph{\bibinfo{title}{{Self-Organized Error Correction in Random Unitary
  Circuits with Measurement}}} (\bibinfo{year}{2020}),
  \eprint{arXiv:2002.12385}.

\end{thebibliography}

\pagestyle{empty}
{ 
\begin{figure*}
\vspace{-1.8cm}
\hspace*{-2cm} 
\includegraphics[page=1]{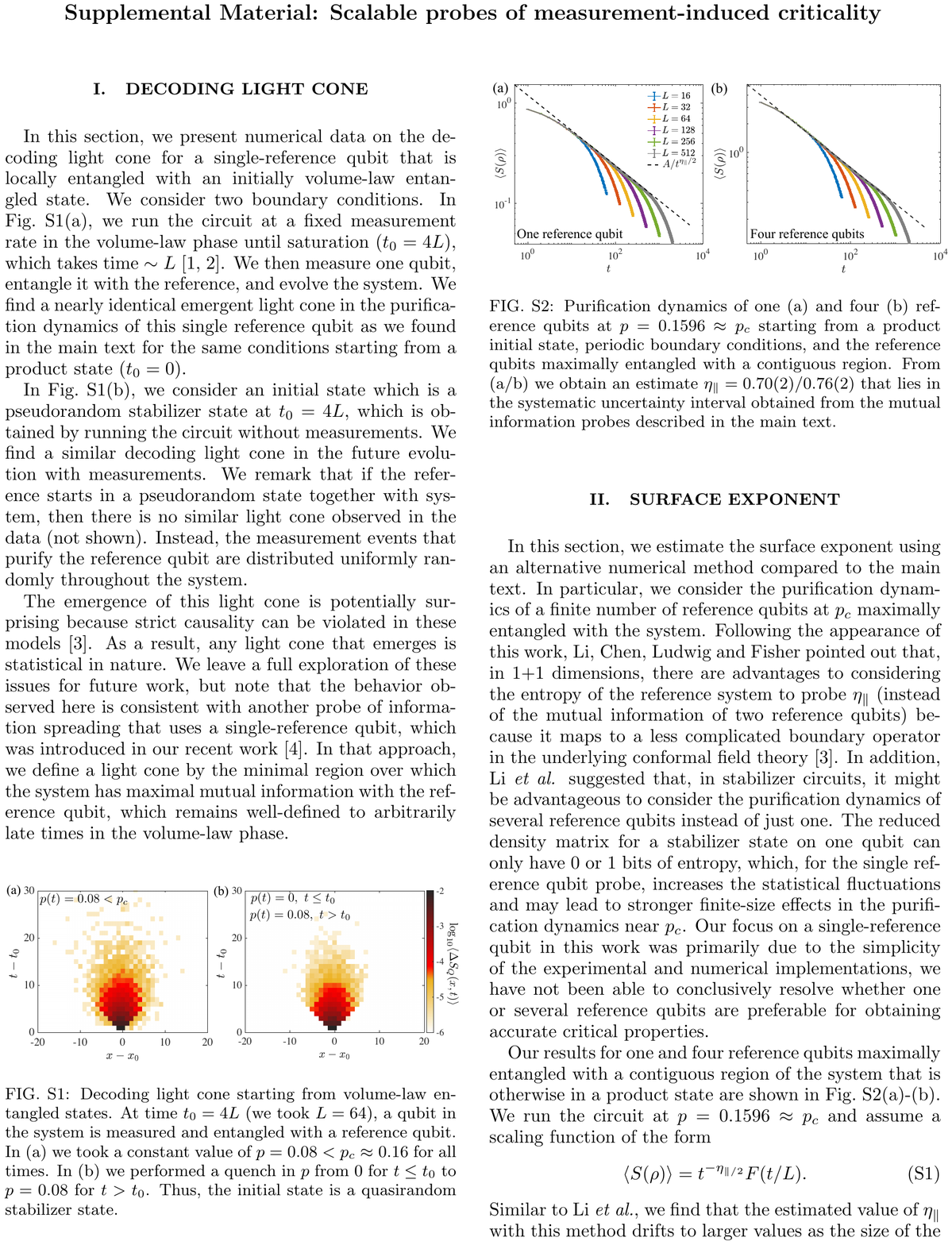}
\end{figure*}

\begin{figure*}
\vspace{-1.8cm}
\hspace*{-2cm} 
\includegraphics[page=2]{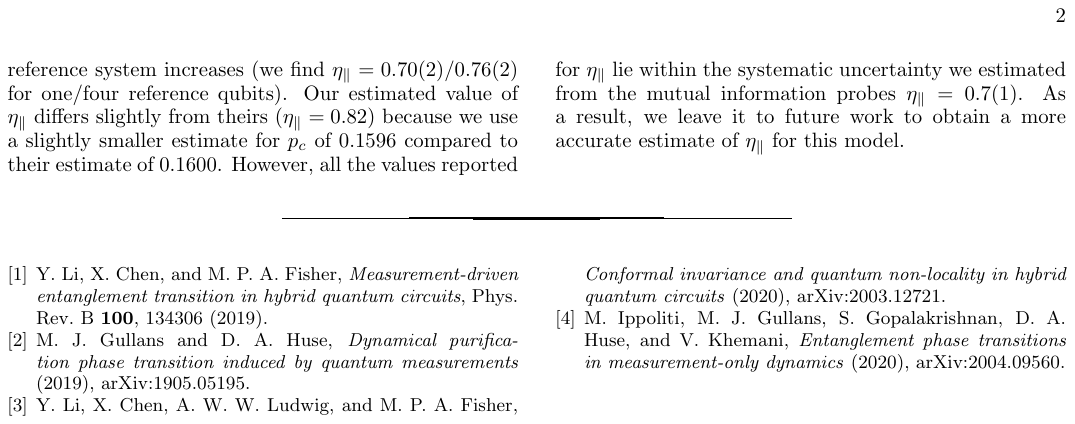}
\end{figure*}

}

\end{document}